\documentclass[%
reprint,
superscriptaddress,
amsmath,amssymb,
prl,
]{revtex4-2}

\usepackage{color}
\usepackage{xcolor}
\usepackage{siunitx}
\usepackage{float}
\usepackage{xspace}
\usepackage{natbib}

\usepackage{graphicx}
\usepackage{dcolumn}
\usepackage{bm}
\usepackage{color}
\usepackage{todonotes} 

\newcommand{\singlecolumn}{86mm}
\newcommand{\doublecolumn}{176mm}

\begin{document}

\title{Enhanced spin coherence while displacing electron in a 2D array of quantum dots}

\author{Pierre-Andr\'e Mortemousque}
\affiliation{Univ. Grenoble Alpes, CNRS, Grenoble INP, Institut N\'eel, 38000 Grenoble, France}%
\affiliation{Univ. Grenoble Alpes, CEA, Leti, F-38000 Grenoble, France}%

\author{Baptiste Jadot}
\affiliation{Univ. Grenoble Alpes, CNRS, Grenoble INP, Institut N\'eel, 38000 Grenoble, France}%
\affiliation{Univ. Grenoble Alpes, CEA, Leti, F-38000 Grenoble, France}%

\author{Emmanuel Chanrion}
\affiliation{Univ. Grenoble Alpes, CNRS, Grenoble INP, Institut N\'eel, 38000 Grenoble, France}%

\author{Vivien Thiney}
\affiliation{Univ. Grenoble Alpes, CNRS, Grenoble INP, Institut N\'eel, 38000 Grenoble, France}%

\author{Christopher B{\"a}uerle}
\affiliation{Univ. Grenoble Alpes, CNRS, Grenoble INP, Institut N\'eel, 38000 Grenoble, France}%

\author{Arne Ludwig}
\affiliation{Lehrstuhl f{\"u}r Angewandte Festk{\"o}rperphysik, Ruhr-Universit{\"a}t Bochum, Universit{\"a}tsstra{\ss}e 150, D-44780 Bochum, Germany}%

\author{Andreas D. Wieck}
\affiliation{Lehrstuhl f{\"u}r Angewandte Festk{\"o}rperphysik, Ruhr-Universit{\"a}t Bochum, Universit{\"a}tsstra{\ss}e 150, D-44780 Bochum, Germany}%

\author{Matias Urdampilleta}
\affiliation{Univ. Grenoble Alpes, CNRS, Grenoble INP, Institut N\'eel, 38000 Grenoble, France}%

\author{Tristan Meunier}
\affiliation{Univ. Grenoble Alpes, CNRS, Grenoble INP, Institut N\'eel, 38000 Grenoble, France}%


\begin{abstract}
The ability to shuttle coherently individual electron spins in arrays of quantum dots is a key procedure for the development of scalable quantum information platforms.
It allows the use of sparsely populated electron spin arrays, envisioned to efficiently tackle the one- and two-qubit gate challenges.
When the electrons are displaced in an array, they are submitted to site-dependent environment interactions such as hyperfine coupling with substrate nuclear spins.
Here, we demonstrate that the electron multi-directional displacement in a $3\times 3$ array of tunnel coupled quantum dots enhances the spin coherence time via the motional narrowing phenomenon.
More specifically, up to 10 configurations are explored by the electrons to study the impact of the displacement on spin dynamics.
An increase of the coherence time by a factor up to 10 is observed in case of fast and repetitive displacement.
The physical mechanism responsible for the loss of coherence induced by displacement is quantitatively captured by a simple model and its implications on spin coherence properties during the electron displacement are discussed in the context of large-scale quantum circuits.
\end{abstract}

\maketitle

The control over the flow of electrons in semiconductor circuits is core to the success of micro and nanoelectronics.
In their quantum counterparts, similar processes are investigated at the single-particle level, to preserve the fragile quantum information stored in individual electron spins \cite{Burkard2002}.
Indeed, displacing these electron spins coherently opens possibilities to convey on-chip quantum information and increase the connectivity of spin-based semiconductor quantum circuits \cite{Taylor2005, Greentree2004, Skinner2003, Barnes2000, Vandersypen2017, Mills2019}.
Several strategies have been recently demonstrated to preserve spin coherence while shuttling charged particles.
A first example consists of the confinement of electrons in moving quantum dots \cite{Bertrand2016, Jadot2020}.
Another protocol is based on repetitive coherent spin tunnelling between adjacent dots.
Such a strategy is particularly appealing for architecture based on large 2D arrays of quantum dots \cite{Vinet2018, Li2018} and has been demonstrated for linear \cite{Fujita2017, Yoneda2020, Ginzel2020} and circular \cite{Flentje2017-2} arrays.
Therefore, understanding the mechanisms at play affecting the fidelity of the displacement in 2D arrays is an important task to optimize this quantum information conveyer procedure and increase as much as possible the distance over which the spin transfer is coherent.

For an electron spin in a quantum dot, the loss of coherence is driven by spin dephasing mechanisms that arise from intrinsic properties of the semiconducting nanostructures like 1/f noise coupled to spin-orbit interaction \cite{Yoneda2018} or hyperfine interaction \cite{Veldhorst2014}.
In both cases, the electron spins are experiencing slow effective magnetic field fluctuations inducing uncertainty on the Larmor precession of individual electron spins.
Therefore, refocusing techniques are extremely efficient on spin systems \cite{Tyryshkin2011, Shulman2014, Wolfowicz2016, Bluhm2011}.
With this property in mind, displacing electron spins on fast timescales is expected to enhance the spin coherence times in virtue of the averaging of the magnetic field fluctuations, as demonstrated in NMR via a phenomenon called motional narrowing effect \cite{Huang2013, Slichter1978}.
The unprecedented level of control recently demonstrated in a 2D array of quantum dots \cite{Mortemousque2020} permits to displace the electron through an important set of possible dot configurations on a timescale faster than the coherence time.
This array turns out to be a relevant platform to investigate and reveal the consequences of electron displacement on spin coherence properties.

Here we study the coherence of individual electron spins when they are displaced via tunnelling through a 2D array of five quantum dots defined in an AlGaAs heterostructure.
In this displacement regime with an electron speed below $100$~\si{\meter\per\second}, the hyperfine interaction with the nuclei of the semiconductor nanostructure is responsible for the loss of coherence \cite{Huang2013}.
This hyperfine interaction results in an effective fluctuating magnetic field, called the Overhauser field, with a distribution imposed by the number of nuclei the electron spins are interacting with.
The larger the number of nuclei, the longer is the coherence time, with a typical square-root dependence \cite{Merkulov2002}.
During the electron displacement, the system visits a number of dot configurations imposed by the gate voltage sequence.
We relate to each dot configuration an effective local magnetic field.
Two principal parameters are influencing the coherence time of the displaced electron spins:
(i) the number of effective local magnetic field configurations explored which can be varied by changing the number of accessible quantum dots in the array, and
(ii) the typical interaction time fixed by the time spent in each charge configuration.
The demonstrated control of the dot system enables us to vary precisely both parameters and to study their impacts on the coherent spin transfer.


\section*{Electron shuttling}

\begin{figure}
\includegraphics[width=\singlecolumn]{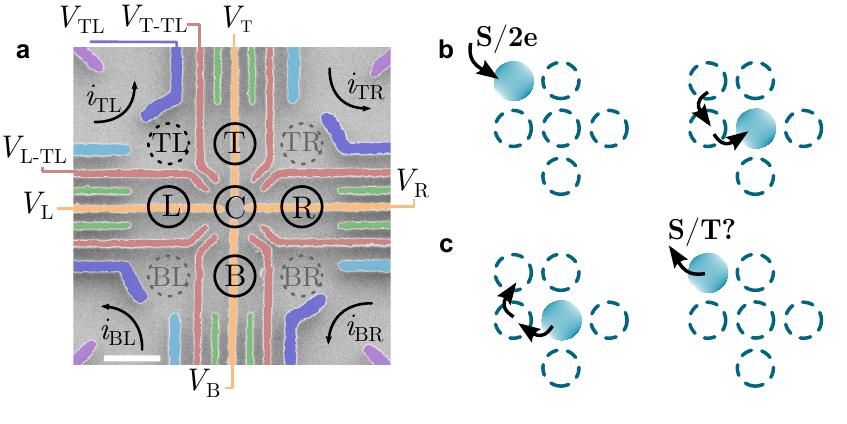}
\caption{
\textbf{Sample and spin loading and readout sequence.}
\textbf{a,}~Electron micrograph of a sample similar to the one used in this work.
The nine circles indicate the $3 \times 3$ array of quantum dots (QDs) arising from the gate-induced potential landscape (see text).
The five dots L, B, R, T, C in a crossbar configuration (solid circles) are used for the coherent electron spin shuttling in 2D.
The square lattice of QDs is electrostatically defined thanks to the \textit{red} gate voltages, and its outer edges by the \textit{green}, \textit{light blue} and \textit{blue} gate voltages.
The couplings of the array to electron reservoirs are achieved through the corner QDs TL, BL, BR, and TR and by controlling the \textit{light blue} and \textit{blue} gates.
The \textit{purple} gates are used to define four quantum point contacts operated as local electrometers, whose conductances set the measured currents $i_\mathrm{TL, BL, BR, TR}$.
\textit{Scale bar} (white) is $200$~\si{\nano\meter}.
\textbf{b,c,}~Schematics of the sequence used to load a singlet spin in C (\textbf{b}), and to read out the final spin state after manipulation (\textbf{c}).
}
\label{fig:FigSample}
\end{figure}

\begin{figure*}[ht]
\centering
\includegraphics[width=\doublecolumn]{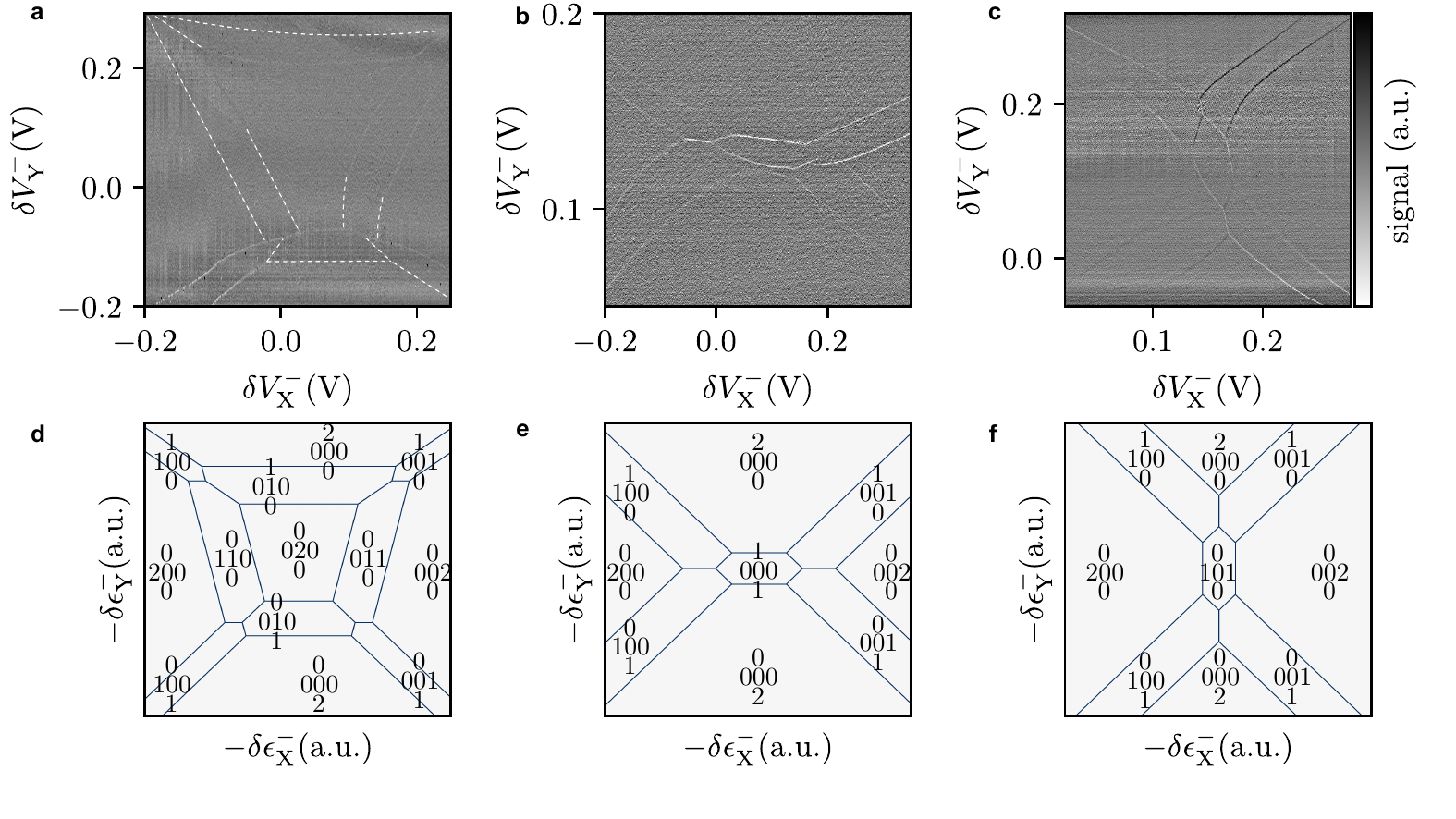}
\caption{
\textbf{Two-electron stability diagrams.}
\textbf{a-c,}~Charge stability diagrams of two electrons in the five dots L, B, R, T, C (\textbf{a}), or in the four dots L, B, R, T (\textbf{b, c}).
The charge state of the array of dots is controlled by sweeping $\delta V_\mathrm{X,Y}^-$.
The signal is recorded as linear combinations of the quantum point contact current derivatives:
(\textbf{a, c}) $\partial_{V_\mathrm{Y}^-} i_\mathrm{TL} + i_\mathrm{BL}$, and
(\textbf{b}) $\partial_{V_\mathrm{Y}^-} i_\mathrm{TL} + i_\mathrm{BL} - i_\mathrm{BR}$.
The dashed white lines in (\textbf{a}) are guide for the eye for weak intensity degeneracy lines.
The stability diagrams are obtained for different gate voltage tunings.
\textbf{d-f,}~Simulations of the stability diagrams (Methods).
The labels indicate the position of the isolated electron in the QD array.
}
\label{fig:FigStabDiag}
\end{figure*}

The sample \cite{Mortemousque2020}, shown in Fig.~\ref{fig:FigSample}a, is an array of $3\times 3$ laterally defined quantum dots (Methods).
The dots are labelled according to their position in the array (e.g. TL stands for Top Left dot).
The system is operated in a mode where the three corner dots TR, BR, and BL are intentionally inaccessible by the electrons (tuned at a higher potential than the crossbar dots L, B, R, T, C).
In this configuration, it is possible to load and isolate two electrons from the reservoir in the TL dot and to transfer them into the dot C by increasing the $V_{\mathrm{L,B,R,T}}$ voltages before rising the potential of the TL dot (Fig.~\ref{fig:FigSample}b).
First, we prove that all the 15 expected charge states for two electrons in five dots ($\binom{5}{1}=5$ where both electrons are in the same dot, $\binom{5}{2}=10$ where electrons are separated in different dots) are accessible by tuning the 4 linear combinations $\delta V^\pm_\mathrm{X,Y}$ of the $V_{\mathrm{L,B,R,T}}$ voltages (Methods).
Figure~\ref{fig:FigStabDiag}a-c shows charge stability diagrams of two electrons recorded as functions of $\delta V^-_\mathrm{X,Y}$, which act like in-plane electric dipole gates, for different values of $\delta V^+_\mathrm{X,Y}$.
By adjusting simultaneously $\delta V^+_\mathrm{X,Y}$, it is possible to change the relative potential of the dot C so that the central region of the diagrams either contains two electrons (Fig.~\ref{fig:FigStabDiag}a), or no electron revealing the regions where the electrons can be separated in (T, B) (Fig.~\ref{fig:FigStabDiag}b) and (L, R) (Fig.~\ref{fig:FigStabDiag}c).
Figure~\ref{fig:FigStabDiag}d-f shows simulations (Methods) of the stability diagrams shown in Fig.~\ref{fig:FigStabDiag}a-c, respectively, in which the identification of the charge distributions is confirmed.
In our precedent report on the coherent control of electron spin in this structure \cite{Mortemousque2020}, we have shown that all charge states exhibited in the stability diagrams of Fig.~\ref{fig:FigStabDiag} can be accessed while setting sufficiently high tunnel couplings for the coherent spin transfers between dots.
Up to 10 charge configurations where the electrons are separated in different dots are explored, and multi-directional and complex one- and two-electron displacements are performed.

\section*{Coherent electron spin shuttling in a 2D array}

We investigate the two-electron spin coherence while the electrons are displaced within the (L, B, R, T, C) subset of quantum dots.
Before all coherent spin shuttling, two electrons in a singlet state (Methods) are loaded in TL before being transferred to L then C (Fig.~\ref{fig:FigSample}b).
Next, a sequence of voltage pulses $\delta V_\mathrm{L, B, R, T}$ is used to separate and displace the electrons coherently in the array.
By shuttling one or two electrons within the dot array, the separated electrons will experience decoherence, resulting in the spin mixing with the triplet states.
Finally, the two electrons are recombined in C and transferred to L and TL for spin readout (Fig.~\ref{fig:FigSample}c).
To consider the three vectorial components of the hyperfine field equally, the experiments are performed under zero external magnetic field \cite{Merkulov2002, Flentje2017-2}.

\begin{figure}
\includegraphics[width=\singlecolumn]{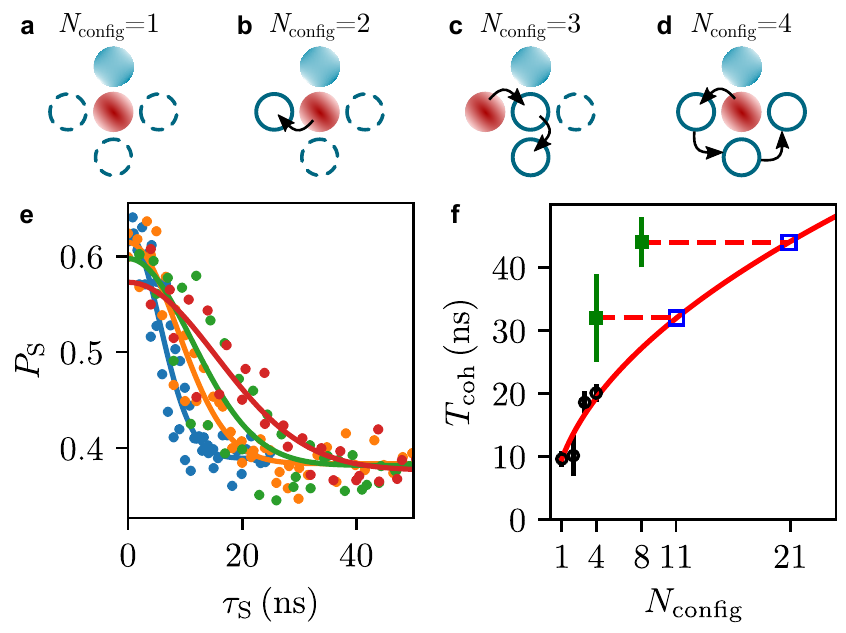}
\caption{
\textbf{Quasi-static electron spin coherence.}
\textbf{a-d,}~Schematic pictures of the different electron spin displacement sequences inside the 2D array of five QDs employed to probe the electron spin coherence shown in (\textbf{e}).
\textbf{e,}~Singlet probability plotted as a function of the total electron separation time $\tau_\mathrm{S}$ spent in one (T-C, \textit{blue}, \textbf{a}), two (T-C and T-L, \textit{orange}, \textbf{b}), three (T-L, T-C, and T-B, \textit{green}, \textbf{c}), and four (T-C, T-L, T-B, and T-R, \textit{red}, \textbf{d}) different charge configurations.
The two electrons are initialised in the singlet state in the C-dot.
Then, one electron is first transferred to T and the second electron visits each other dot only once over an equal time of $\tau_R$.
As a consequence, the second electron spends for the different sequences a total separation time $\tau_S = 1\times \tau_R$ in C,  $\tau_S = 2\times \tau_R$ in C and L, $\tau_S = 3\times \tau_R$ in L, C, and B, or $\tau_S = 4\times \tau_R$ in C, L, B, and R, where $\tau_R$ is in integer value of the AWG clock period.
The data are fitted with Gaussian decays (\textit{solid lines}) with characteristic times $T_2^\star$ equal to $7.9\pm0.6$ (blue), $12.3\pm0.7$ (orange), $15.8\pm1.5$ (green), and $21.9\pm1.6$~\si{\nano\second} (red).
\textbf{f,}~$\overline{T_2^\star}$ averaged over different possible charge configurations plotted as a function $N_\textrm{config}$ (black circles).
The data are fitted with a square-root function (red solid line) $\overline{T_2^\star(N_\mathrm{config}=1)} \sqrt{N_\mathrm{config}}$, with $\overline{T_2^\star(N_\mathrm{config}=1)}=9.6(0.7)$.
The experimental data of Fig.~\ref{fig:FigExpSimDisplacement}c,d are plotted as green squares ($T_\textrm{coh}$ of $32\pm7$ and $44\pm4$~\si{\nano\second}, respectively), and their projection on the square root curve as open blue squares (equivalent $N_\textrm{config}$ of $11$ and $21$, respectively).
}
\label{fig:FigStaticElectrons}
\end{figure}

We first investigate the quasi-static case where the electrons are separated in different dots, one electron remaining in one dot, the other visiting a single time an increasing number of dots.
For a first electron fixed in dot T, it results in $N_\textrm{config}=1$ to 4 distinct possible configurations (Fig.~\ref{fig:FigStaticElectrons}a-d, respectively).
The corresponding voltage pulse sequences are depicted in Supp. Fig.~\ref{supp:fig:static_sequences}.
For each experiment, the time spent in each configuration, referred to as the resting time $\tau_R$, is varied.
The resulting singlet probabilities are plotted (filled circles) in Fig.~\ref{fig:FigStaticElectrons}e as a function of the total separation time $\tau_S = N_\textrm{config} \times \tau_R$.
The coherence time is directly extracted by fitting a Gaussian decay $\textrm{e}^{-(\tau_S/T_\textrm{coh})^2}$ (solid lines) and increases with $\sqrt{N_\mathrm{config}}$.
Indeed, the separated electrons explore a large ensemble of nuclear spins during the displacement, averaging the hyperfine interaction \cite{Flentje2017, Merkulov2002, Bloembergen1948}.
We have repeated this set of experiments for different permutations of QDs to extract a statistical dependence of $T_\textrm{coh}$ on $N_\textrm{config}$ and average out the influence of the non-regular coupling between the different dots \cite{Mortemousque2020}.
For each size of visited subsets, the average and the standard deviation of $T_\textrm{coh}$ is plotted in Fig.~\ref{fig:FigStaticElectrons}f as a function of $N_\textrm{config}$.
They quantitatively fit with the expected square root law (solid line).
It confirms the importance of the hyperfine interaction and the number of nuclei the electrons interact with.
It also demonstrates the successful coherent displacement within the 2D array of dots.


\section*{Enhancement of the spin coherence time via electron shuttling}

\begin{figure}[!t]
\includegraphics[width=\singlecolumn]{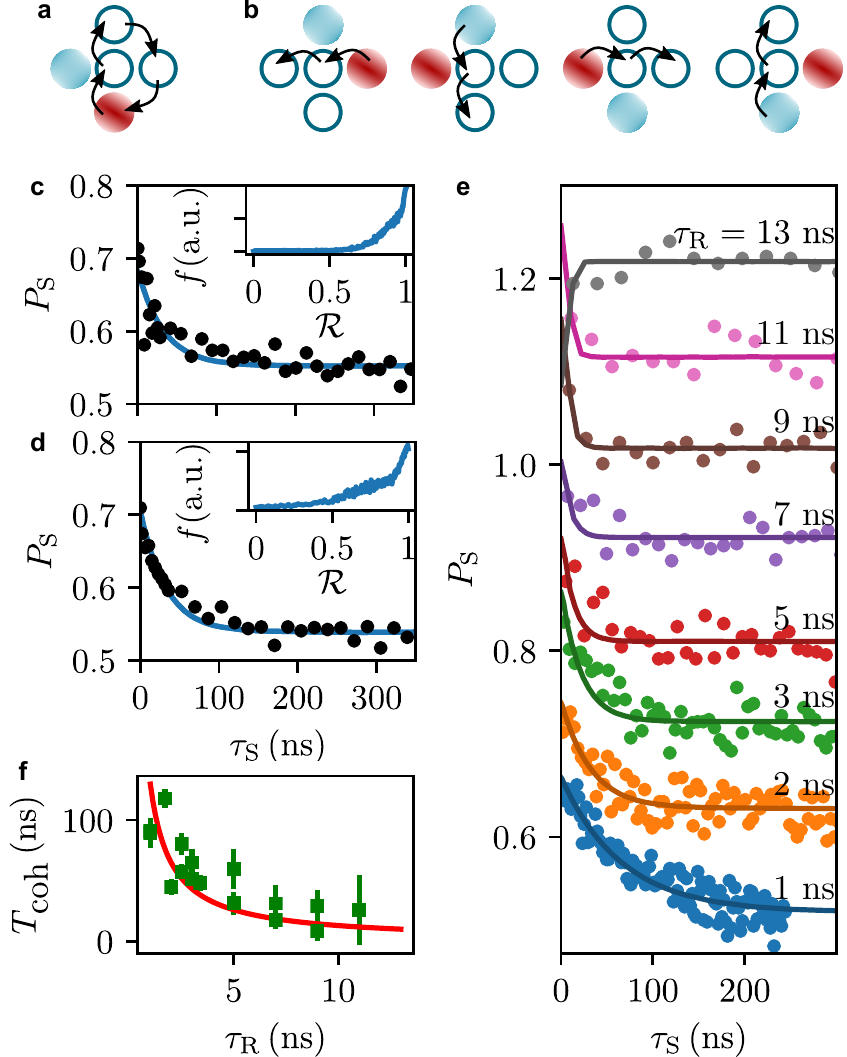}
\caption{
\textbf{Enhanced spin coherence time for periodic electron displacements.}
\textbf{a-b,}~Schematic pictures of the different electron spin periodic displacement sequences inside the 2D array of five QDs employed to probe the electron spin coherence shown in (\textbf{c-f}).
\textbf{c, d,}~Singlet probabilities plotted (black dots) as a functions of the total electron separation time $\tau_\mathrm{S}$, \textbf{c} for the single electron periodic displacement pattern (\textbf{a}), and \textbf{d} for the two-electron periodic displacement pattern (\textbf{b}).
The data were acquired for configuration durations $\tau_\mathrm{R}$ of $1.7$~\si{\nano\second} (\textbf{c}) and $2.1$~\si{\nano\second} (\textbf{d}).
The experimental data are fitted using simulations of the spin dynamics (see text) as solid blue lines.
The fit error functions (see text) are shown in insets.
\textbf{e,}~Singlet probability plotted (dots) as a function of the total electron separation time $\tau_\mathrm{S}$, for increasing (bottom to top) values of $\tau_R$: 1, 2, 3, 5, 7, 9, 11, and 13~\si{\nano\second}, for the two-electron periodic displacement case (\textbf{b}).
The solid lines are the simulated singlet probabilities (see text) computed for the corresponding $\tau_R$ times, for a common remanence value $\mathcal{R}=0$.
\textbf{f,}~Characteristic coherence times $T_\textrm{coh}$ plotted as functions of $\tau_R$ and fitted with an inverse function  $T_\textrm{coh}^0/\tau_R+T_2^\star$, with $T_\textrm{coh}^0 = 130\pm24$~\si{\nano\second}.
}
\label{fig:FigExpSimDisplacement}
\end{figure}

We now analyse the impact of the electron tunnelling on their spin coherence.
In comparison with the previous case where the electrons explore a single time distinct charge configurations, we study the situation where the charge configurations are periodically explored.
For each experiment, the resting time spent in each charge configuration $\tau_R$ is set to a constant value.
We use two periodic pulse sequences corresponding to two-electron displacement patterns: (i) single- (ii) two-electron periodic displacement.
In the single-electron periodic displacement case, the first electron is maintained in L, and the second is displaced among B, C, T, and R, so that four isochronous dot configurations are periodically visited (Fig.~\ref{fig:FigExpSimDisplacement}a and Supp. Fig.~\ref{supp:fig:displacement_sequences}a).
In the two-electron periodic displacement case, an electron is first moved in B and the second in R.
From this point, each electron is sequentially displaced along a two-step trajectory, either vertically, or horizontally.
Eight isochronous QD configurations are periodically visited with C the only dot visited by both electrons (Fig.~\ref{fig:FigExpSimDisplacement}b and Supp. Fig.~\ref{supp:fig:displacement_sequences}b).
The experimental singlet probabilities (black points) recorded for the single-electron ($\tau_R=1.7$~\si{\nano\second}), and for the two-electron ($\tau_R=2.1$~\si{\nano\second}), are plotted in Fig.~\ref{fig:FigExpSimDisplacement}c,d, respectively.
The curves are fitted by exponential decays (Supp. Figure~\ref{supp:fig:exponential_displaced}a,b), with characteristic times of $32\pm7$ and $44\pm4$~\si{\nano\second}, corresponding to coherence length of $1.9\pm0.4$ and $2.2\pm0.2$~\si{\micro\meter} (assuming a conservative interdot distance of 100~\si{\nano\meter}), respectively.
These decay times are plotted as green squares in Fig.~\ref{fig:FigStaticElectrons}f, and their projection on the square root curve as open blue squares ($N_\textrm{config}$ equivalent to $11$ and $21$, respectively).
A significant improvement is observed compared to the quasi-static case: they are 1.6 times higher than the expected coherence time for four and eight static dot configurations, respectively.
Moreover, the decoherence law changes from Gaussian-like (Fig.~\ref{fig:FigStaticElectrons}e) to exponential (Fig.~\ref{fig:FigExpSimDisplacement}c,d) decay when visiting the charge configuration multiple times, which confirms the significant impact of the electron dynamics on their spin coherence.

By varying the resting time $\tau_R$, we can explore different shuttling regimes for the two-electron periodic displacement case.
The resulting singlet probability is plotted in Fig.~\ref{fig:FigExpSimDisplacement}e as a function of $\tau_S = (8N_\textrm{cycle}+1)\tau_R$ for different values of $\tau_R$ going from 1~\si{\nano\second} to 13~\si{\nano\second} (bottom to top).
No singlet probability decay is observed for $\tau_R$ longer than the static spin coherence time due to complete mixing between the singlet and triplet states.
By progressively reducing $\tau_R$, we observe the emergence of a single exponential decay of the singlet probability, demonstrating preservation of the spin coherence during displacement.
We extract the characteristic decay time $T_\textrm{coh}$ by fitting with exponential decay (Supp. Fig.~\ref{supp:fig:exponential_displaced}c) which we plot against $\tau_R$ (Fig.~\ref{fig:FigExpSimDisplacement}f).
The points are fitted with an inverse function $\alpha / \tau_R$, typical of a motional narrowing phenomenon, with a scale factor $\alpha=130\pm24$~\si{\nano\second^2}.
This value is in agreement with the expected square of the dephasing time \cite{Slichter1978}, estimated in Fig.~\ref{fig:FigStaticElectrons}f to be  $\overline{T_2^\star(N_\mathrm{config}=1)}^2=9.6^2=93(13)$~\si{\nano\second^2}.

\begin{figure}
\includegraphics[width=\singlecolumn]{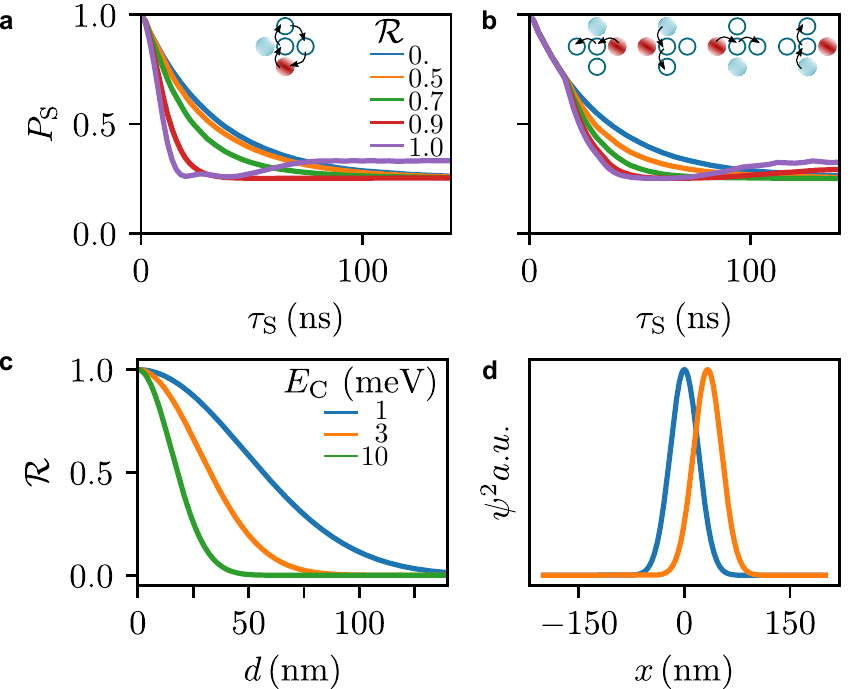}
\caption{
\textbf{Electron path randomisation consequences.}
\textbf{a,b,}~Simulated decoherence decay curves for one- (\textbf{a}) and two- (\textbf{b}) displaced electrons for different values of the remanence parameter $\mathcal{R}$: 0 (blue), 0.5 (orange), 0.7 (green), 0.9 (red), and 1 (purple).
\textbf{c,}~Estimated remanence $\mathcal{R}$ plotted as a function of the distance between the dot location of two consecutive periods, for a charging energy of 1 (blue), 3 (orange), and 10~\si{\milli\electronvolt} (green).
\textbf{d,}~Cross-sectional view of two-electron densities computed using the fundamental Fock-Darwin state \cite{Kouwenhoven2001}, and distant of 33~\si{\nano\meter} ($\mathcal{R}=0.5$), assuming a charging energy of 3~\si{\milli\electronvolt}.
}
\label{fig:FigDynamics}
\end{figure}


\section*{Path randomisation}

If only a definite set of dot configurations are cyclically explored, no difference in the spin dynamics should be observed between the quasi-static and the periodic displacement protocols.
Then, the only relevant parameter is the time spent in each configuration, and a gaussian decay of the singlet probabilities would be expected.
However, the difference observed in the experiment forces us to reconsider the initial assumptions made about the electron displacement:
the electrons are not periodically exploring the \textit{same} set of random magnetic field configurations.
To better understand the impact of the electron shuttling on spin coherence time, we model the system and study the spin dynamics (Supp. Fig.~\ref{supp:fig:simulation_model} and Methods).
We introduce the notion of randomization of the effective hyperfine field during the coherent electron displacements.
It takes the form of a remanent parameter $0 \leq \mathcal{R} \leq 1$, and it describes the change in effective hyperfine field, between two consecutive displacement periods $i$ and $i+1$, at equivalent dot $j$.
It is expressed as $\vec{B}_{i+1}^j=\mathcal{R}\vec{B}_{i}^j+(1-\mathcal{R})\vec{\delta B}_{i+1}^j$, where $\vec{\delta B}_{i+1}^j$ is a random occurence of the magnetic field distribution.
For example, a full remanence $\mathcal{R}=1$ means that the effective hyperfine interaction for an electron visiting multiple times the same dot is constant over the entire displacement sequence of a single shot.
Assuming (i) the evolution of the effective hyperfine field much slower than the coherent spin displacement \cite{Reilly2008, Nakajima2020}, (ii) the nuclear spin immune to the electrostatic manipulation, and (iii) not affected by the electron motion, the remanence value is then interpreted as a small electrostatic potential reorganization, leading to fluctuations in the dot positions.
For $\mathcal{R}<0.5$, the decoherence decay curves are exponential-like (see Fig.~\ref{fig:FigDynamics}a,b) and only weakly dependent on $\mathcal{R}$, which is the fingerprint of a continuous motional narrowing effect.
On the other side, for $\mathcal{R}>0.5$, the motional narrowing process is limited by the number of visited sites, which leads to Gaussian or Gaussian-like \cite{Merkulov2002} coherence decays.
Such a transition from Gaussian to exponential dynamics is a clear manifestation of the motional narrowing effect that has been observed in liquid NMR.

The simulation well reproduces the experimental data in the single- and two-electron displacement with a remanence parameter $\mathcal{R}\approx 0$ (solid blue curve in Fig.~\ref{fig:FigExpSimDisplacement}c,d).
The evaluation of the best fitting simulated coherence decay curves (calculated to minimize the sum of the square differences with the experimental data, insets) show that the data of the two experiments can be reproduced with a value of remanence $0\lesssim\mathcal{R}<0.5$.
One can note that the difference in decay times between the single- and the two-electron displacement cases is not significant regarding the fit uncertainties (Supp. Figure~\ref{supp:fig:exponential_displaced}a,b).
Moreover, this parameter also provides simulated coherence decays very similar to the experimental data of Fig.~\ref{fig:FigExpSimDisplacement}e, for the various values $\tau_R$ of the displacement dynamics (solid lines), as well as for the data obtained in the quasi-static experiments (Supp. Fig.~\ref{supp:fig:gaussian_quasistatic}).

Therefore, the cyclic nature of the evolution has to be randomized along the path of the electrons to reproduce the experimental data.
Since the hyperfine interaction is a contact interaction, a displacement of the size of the dot will be sufficient to change the effective magnetic field completely.
The relationship between $\mathcal{R}$ and the electron centre of mass separation between two consecutive periods $d$ is given for various charging energies in Fig.~\ref{fig:FigDynamics}c.
It is calculated as the overlap between the electron density at two consecutive periods $\mathcal{R} = \int \psi^2_i \psi^2_{i+1}dr$.
Therefore, between two consecutive displacement periods, the electron centre of mass is expected to be distant of about $d\approx33$~\si{\nano\meter} for a charging energy of 3~\si{\milli\electronvolt} (Fig.~\ref{fig:FigDynamics}d and Supp. Fig.~\ref{supp:fig:shallowpotential}).
The schematic electron centre of mass trajectories in the single- and the two-electron displacement experiments are superimposed on the gate pattern and the simulated electrostatic potential of Supp. Fig.~\ref{supp:fig:shallowpotential}a and b, respectively.

Here we speculate that these fluctuations in dot positions may occur because of the large voltage ranges employed to displace the electron through the quantum dot array (as large as few hundreds of millivolts, Supp. Fig.~\ref{supp:fig:displacement_sequences}).
Such energy is expected to induce perturbations of the semiconductor nanostructure resulting in a change in the disorder potential.
Besides, the electrostatic potential simulations of Supp. Fig.~\ref{supp:fig:shallowpotential}a,b confirm that the confinement potentials remain shallow during the electron shuttling, making the dot positions very sensitive to fluctuations of charges in the substrate (e.g. donors in the doping layer of the heterostructure, Supp. Fig.~\ref{supp:fig:shallowpotential}c).
Moreover, the typical timescale of this alteration is comparable to the one of the excitation potential and therefore can be as fast as 1 nanosecond.
It implies that the electrons are displaced along a path that is fluctuating on a timescale comparable to the time needed to explore few dot configurations.
As a consequence, the effective magnetic field configuration is not precisely cyclic anymore and become randomized.

A similar assumption on the electron path has been reported recently by our group in a different displacement regime \cite{Jadot2020}.
The electrons were then displaced in moving quantum dots at a speed of 3000~\si{m/s}.
Indeed, at this speed, the motional narrowing process is extremely efficient for hyperfine interaction, and the main mechanism for decoherence is mediated by spin-orbit interaction.
Evidence of change in the disorder potential was imprinted in the coherence of the transported electrons initially prepared in a superposition of antiparallel spin states.


\section*{Conclusions}
We have performed the one- and two-electron coherent shuttling in a two-dimensional array and explored various displacement trajectories through different sets of quantum dots.
We report an increase of the coherence time, which corresponds to a coherence length well above one micron for both one- and two-electron displacements.
Furthermore, we specifically study the zero magnetic field regime where all the three components of the hyperfine interaction contribute equally to the decoherence of the two electron-spin singlet states.
This hyperfine interaction is expected to be averaged during the electron displacement.
A signature of a motional narrowing process is observed with a decoherence rate inversely proportional to the time spent in the static phase.
Indeed, reducing the time where the electrons are effectively static and averaging faster than the spin dynamics over many nuclear spin configurations increase the observed coherence time \cite{Huang2013}.
An important consequence of the observation is the signature that the path of the electron is slightly modified from one period of displacement to the next.
It results in the randomization of the effective field distribution experienced by the electrons.
Experimental results are reproduced with a simple model.
Therefore, only the single electron displacement during the gate movement from one dot to another is relevant to explain the increase in spin coherence time.



\section*{Methods}

\noindent The methods about the spin initialisation in the singlet state, the electrostatic potential, the stability diagram simulations and the virtual gates employed in this study are essentially similar to \cite{Mortemousque2020}.

\noindent\textbf{Acquisition protocol.}
The voltage control apparatus is programmed to execute a list of single-shot experiments (about a few thousand different sequences).
The total duration of each single-shot (comprising the electron spin initialisation, manipulation, readout and the instrumental over-head) is about 50~\si{\milli\second}.
The singlet spin probabilities are computed by averaging 1000 single-shot (repetition of the programmed list of sequences).
Therefore, the typical time between the repetition of two equivalent sequences is about a few minutes.

\noindent\textbf{Materials and set-up.}
Our device was fabricated using a Si-doped AlGaAs/GaAs heterostructure grown by molecular beam epitaxy, with a two-dimensional electron gas 100~\si{\nano \meter} below the crystal surface which has a carrier mobility of $10^6$~\si{cm^2V^{-1}s^{-1}} and an electron density of $2.7\times10^{11}$~\si{cm^2}.
The quantum dots are defined using electrostatic confinement generated using Ti/Au Schottky gates.
It is anchored to the cold finger, which is in turn mechanically attached to the mixing chamber of a homemade dilution refrigerator with a base temperature of $60$~\si{mK}.
It is placed at the centre of a superconducting solenoid generating the static out-of-plane magnetic field.
Quantum dots are defined and controlled by applying negative voltages on Ti/Au Schottky gates deposited on the surface of the crystal.
Homemade electronics ensure fast changes of both chemical potentials and tunnel couplings with voltage pulse rise times approaching 100~\si{\nano\second} and refreshed every 16~\si{\micro\second}.

A Tektronix 5014C with a typical channel voltage rise time (20\% - 80\%) of $0.9$~\si{\nano\second} are used to rapidly change the $V_\mathrm{L}$, $V_\mathrm{B}$, $V_\mathrm{R}$, and $V_\mathrm{T}$ gate voltages.
The charge configurations can be read out by four quantum point contacts, tuned to be sensitive local electrometers, and independently biased with 300~\si{\micro\volt}.
The resulting currents $i_\text{TL}$, $i_\text{BL}$, $i_\text{BR}$ and $i_\text{TR}$ are measured using current-to-voltage converters with a typical bandwidth of 10~\si{\kilo\hertz}.
The single-shot repetition rate is about 20 Hz.

\noindent\textbf{Nuclear spin dynamics and experimental timescales}
The experimental singlet probabilities are calculated as the average spin readout results of either 150 (Fig.~\ref{fig:FigDynamics}a), or single 1000 shots (Figs.~\ref{fig:FigStaticElectrons}b and \ref{fig:FigExpSimDisplacement}c,d).
Sequences of multiple displacement patterns are repeated to get a large enough number of shots.
A sequence has a typical duration of 1 day, with a sequence length of $\lesssim 4000$ different shots, and a shot duration of about 50~\si{\milli\second}.
Therefore, the time between two equivalent shots is about 200~\si{\second}, which is much shorter than the nuclear spin relaxation time in the dipole-dipole field of other nuclear spins is in the range of 100~\si{\micro\second} \cite{Merkulov2002, Malinowski2017}.
It allows the nuclear spin distribution to be randomised between each equivalent single-shot experiment.
Consequently, there is no preferential quantification axis at zero magnetic fields, and the spin-orbit interactions are averaged out.

\noindent\textbf{Simulation of the electron spin dynamics.}
The simulations are performed with an initial singlet spin state.
A four-level system is considered to model the 2 electron spins, each submitted to a specific magnetic field (random hyperfine interaction with the substrate nuclear spins).
The numerical time-integration of the Schr{\"o}dinger equation is computed for 1000 different couples of magnetic field vectors, following a centered Gaussian distribution with a standard deviation of 3.9~\si{\milli\tesla} (see Fig.~\ref{supp:fig:simulation_model}).
Magnetic field remanence parameters are added to the model to consider the spatial evolution of the effective dot locations during the different displacement patterns.
Finally, the average singlet probability is calculated \cite{Johansson2012, Johansson2013}.


\bibliography{biblio}

\section*{Acknowledgements}

We would like to thank Maud Vinet, Xuedong Hu, and Lieven M. K. Vandersypen for enlightening discussions.
We acknowledge technical support from the Poles of the Institut N\'eel, and in particular, the NANOFAB team who helped with the sample realisation, as well as E. Eyraud, T. Crozes, P. Perrier, G. Pont, H. Rodenas, D. Lepoittevin, C. Hoarau and C. Guttin.
M.U. acknowledges the support of project CODAQ (ANR-16-ACHN-0029).
A.L. and A.D.W. acknowledge gratefully the support of
DFG-TRR160,
BMBF-Q.com-H 16KIS0109,
and the DFH/UFA CDFA-05-06.
T.M. acknowledges ﬁnancial support from ERC QSPINMOTION, ERC QUCUBE, ANR CMOSQSPIN (Grant No. ANR-17-CE24-0009), ANR SiQuBus and UGA IDEX (Grant No. ANR-15-IDEX-02).

\section*{Author contributions statement}
P.-A. M. fabricated the sample and performed the experiments with the help of T.M. and C.B.. P.-A. M. and T.M. interpreted the data. P.-A. M. and T.M. wrote the manuscript with the input of all the other authors.
H.F. contributed to the experimental setup.
A.L. and A.D.W. performed the design and molecular-beam-epitaxy growth of the high mobility heterostructure.
All authors discussed the results extensively, as well as the manuscript.

\textbf{Additional information}
Correspondence and requests for materials should be addressed to
tristan.meunier@neel.cnrs.fr or
pierre-andre.mortemousque@cea.fr

\makeatletter
\widetext
\clearpage
\begin{center}
\textbf{\large Supplemental Materials: \@title}
\end{center}
\setcounter{equation}{0}
\setcounter{figure}{0}
\setcounter{table}{0}
\setcounter{page}{1}
\renewcommand{\theequation}{S\arabic{equation}}
\renewcommand{\thefigure}{S\arabic{figure}}
\renewcommand{\bibnumfmt}[1]{[#1]}
\renewcommand{\citenumfont}[1]{#1}

\begin{figure*}[!h]
\includegraphics[width=\doublecolumn]{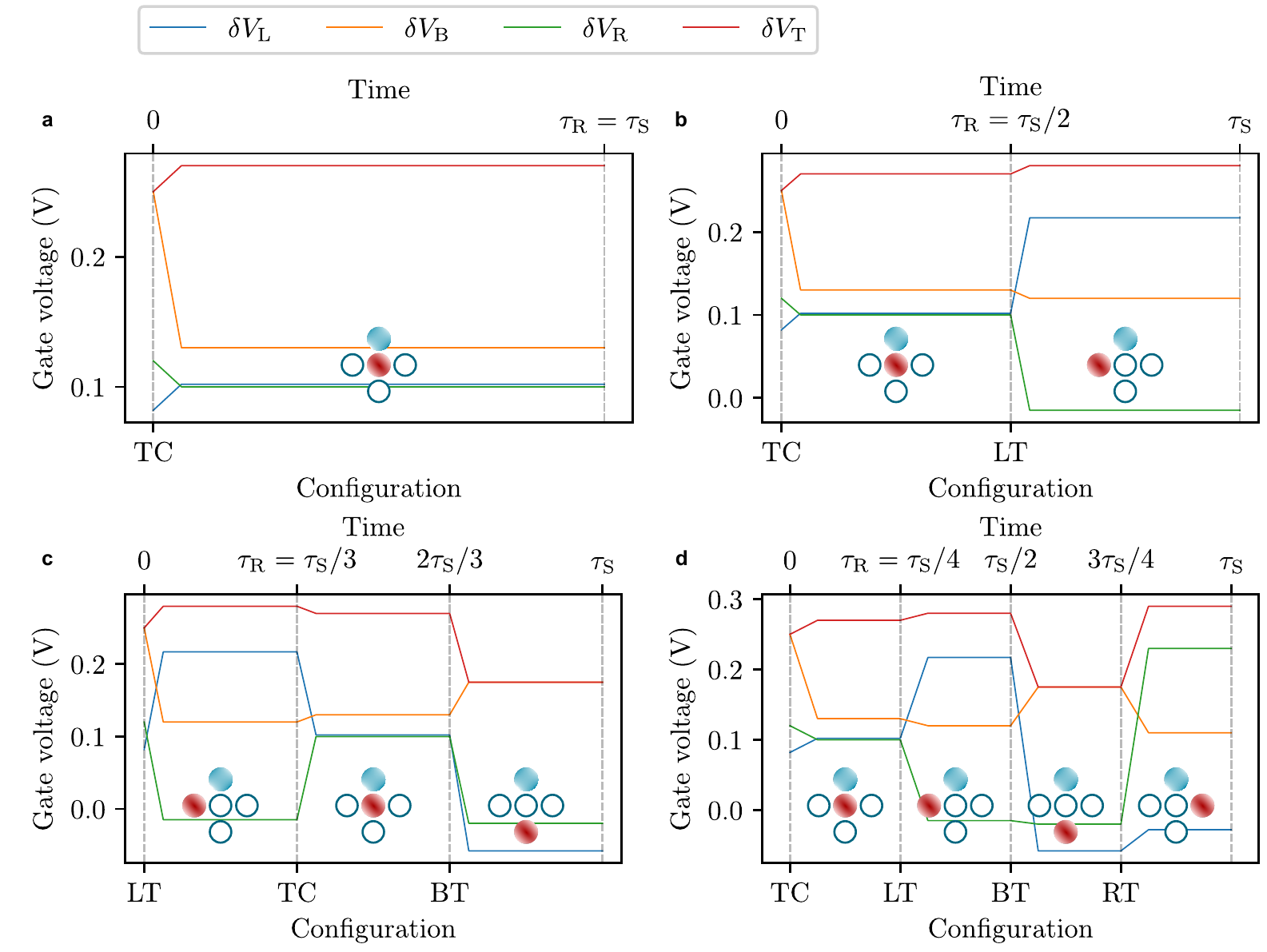}
\caption{
\label{supp:fig:static_sequences}
\textbf{Single-period electron displacement sequences.}
\textbf{a-d,}~Sequences of nanosecond voltage pulses used to perform the single period electron displacement of Fig.~\ref{fig:FigStaticElectrons}.
\textbf{a,}~The pulses are applied during $\tau_S$ to displace the electrons in the TC configuration.
\textbf{b,}~The pulses are applied during $\tau_S/2$ to displace the electrons in the TC and LT configurations.
\textbf{c,}~The pulses are applied during $\tau_S/3$ to displace the electrons in the LT, TC and BT configurations.
\textbf{d,}~The pulses are applied during $\tau_S/4$ to displace the electrons in the TC, LT, BT and RC configurations.
The x-axes labels denote the location of the dot occupied by the electron.
For example, TC stands for one electron in dot T, and one in dot C.
The initial and final configurations correspond to the two electrons in the C dot.
The schematic representations of the charge configurations are superimposed.
Only one period of displacement is shown for each sequence.
The voltage rising times are given as illustrations.
}
\end{figure*}

\clearpage
\begin{figure*}[!h]
\includegraphics[width=\doublecolumn]{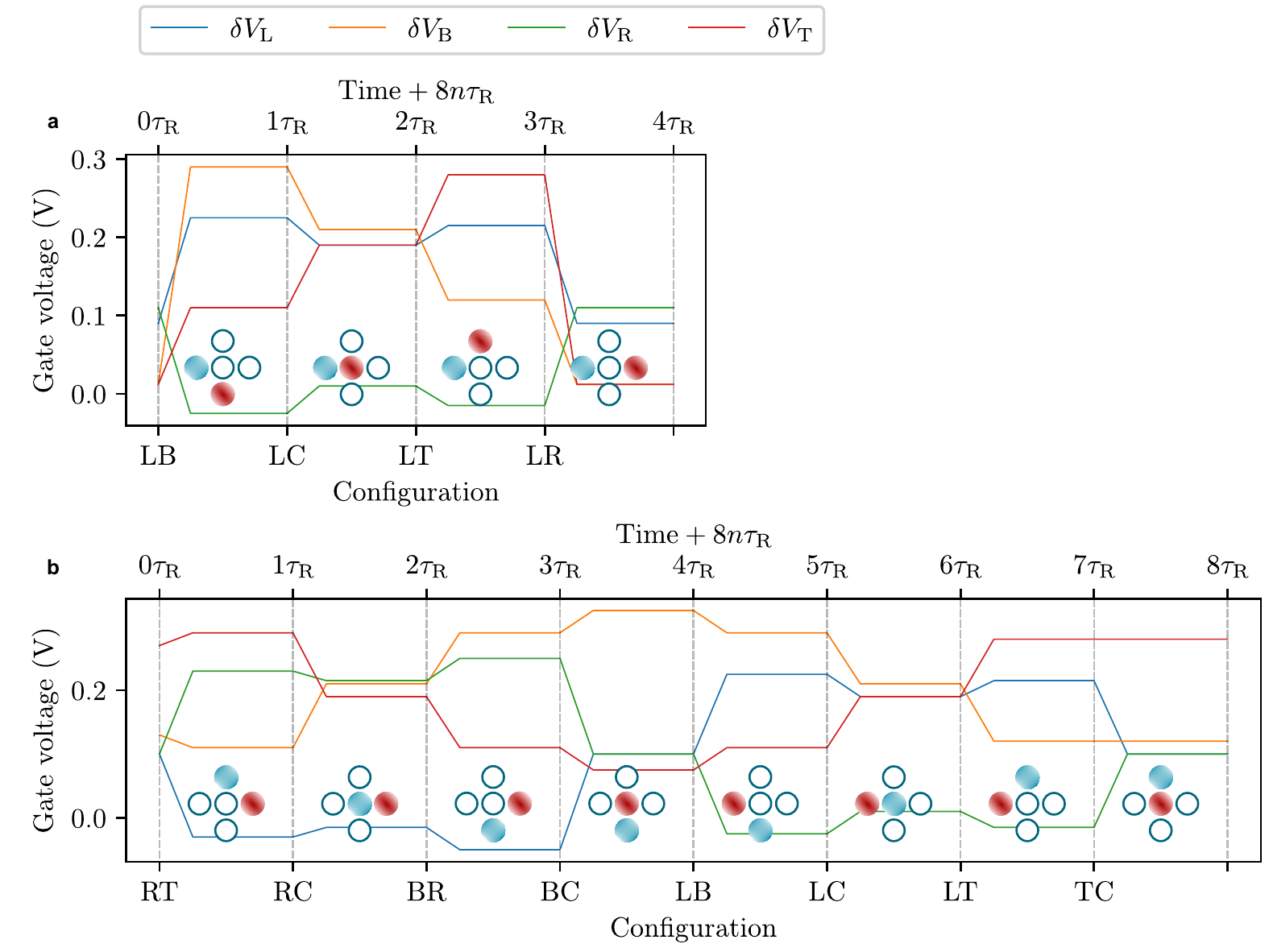}
\caption{
\label{supp:fig:displacement_sequences}
\textbf{Periodic electron displacement sequences.}
\textbf{a,b,}~Sequences of nanosecond voltage pulses used to perform the periodic electron displacements.
\textbf{a,}~One electron remains in the left quantum dot while the second electron is displaced in the B, C, T, R dots.
\textbf{b,}~One electron periodically visits L, C, R, and the second T, C, B.
The x-axes labels denote the location of the dot occupied by the electron.
For example, LB stands for one electron in dot L, and one in dot B.
The schematic representations of the charge configurations are superimposed.
Only one period of displacement is shown for each sequence.
The voltage rising times are given as illustrations.
}
\end{figure*}

\clearpage
\begin{figure}[!h]
\includegraphics[width=\singlecolumn]{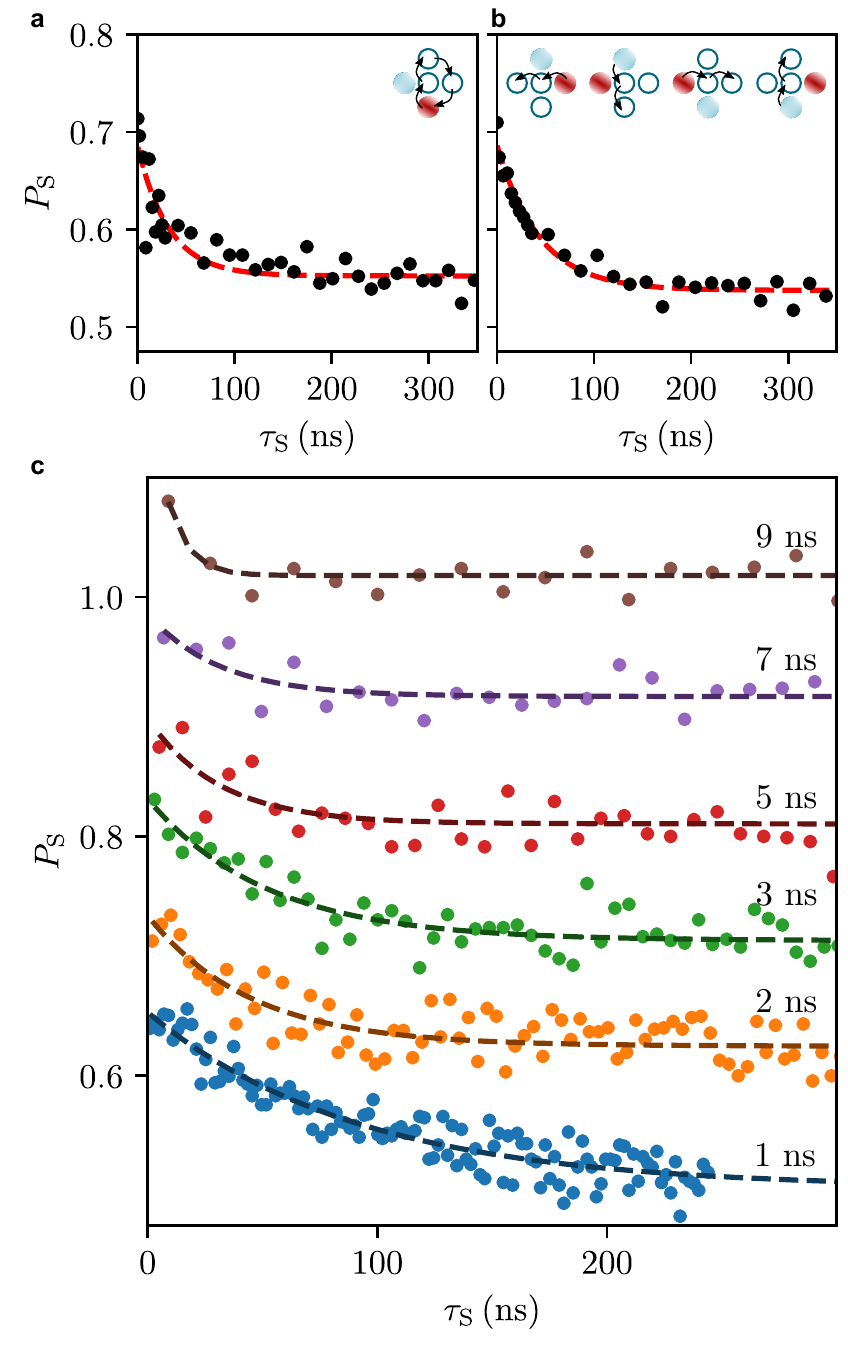}
\caption{
\textbf{Exponential spin coherence fits of the periodically displaced electron.}
\label{supp:fig:exponential_displaced}
In order to extract the coherence time $\textrm{T}_\mathrm{coh}$ that are discussed in the manuscript, we fit the experimental data using a single exponential decay law.
\textbf{a, b}~Same dataset as in Fig.~\ref{fig:FigExpSimDisplacement}c and d, respectively.
The data are fitted with an exponential decay (\textit{solid lines}) with characteristic times of  (\textbf{a}) $32\pm7$, and (\textbf{b}) $44\pm4$~\si{\nano\second} (red).
\textbf{c}~Same dataset as in Fig.~\ref{fig:FigExpSimDisplacement}e.
The data are fitted with an exponential decay (\textit{solid lines}) with the characteristic times:
$91.3\pm10$ (blue),
$44.8\pm7$ (orange),
$51.7\pm8$ (green),
$31.3\pm9$ (red),
$31.2\pm15$ (purple), and
$8.9\pm9$ (brown).
}
\end{figure}


\clearpage
\begin{figure}[!h]
\includegraphics[width=\singlecolumn]{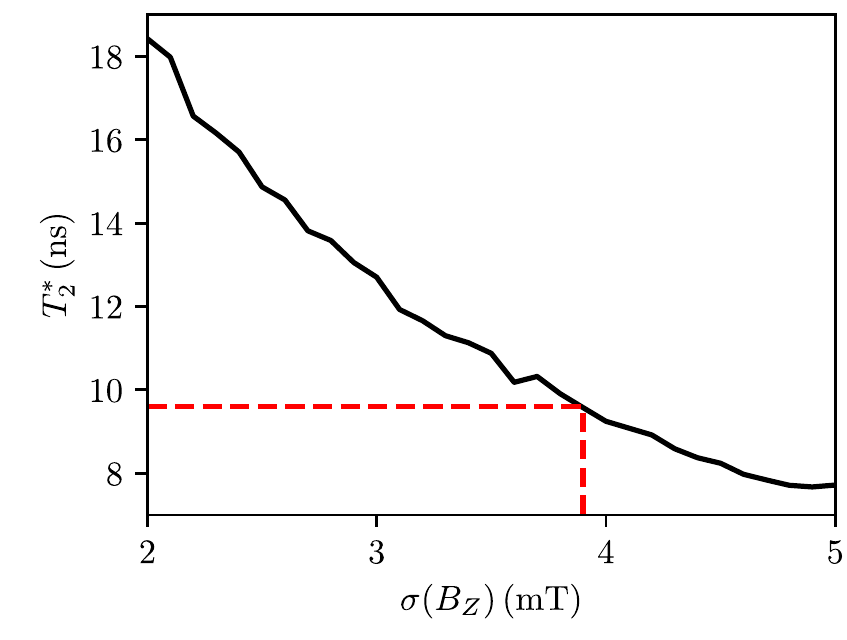}
\caption{
\label{supp:fig:simulation_model}
\textbf{Simulated spin coherence time in a static double quantum dot.}
The electron spin decoherence curves can be simulated for different standard deviation $\sigma(B_\textrm{Z})$ values of a Gaussian distribution in hyperfine fields (Methods).
The coherence times extracted from Gaussian fits are plotted as functions of $\sigma(B_\textrm{Z})$.
The field corresponding to the spin coherence time $T_2^\star(N_\textrm{config}=1)=9.6$~\si{\nano\second} (extracted from the $T_2^\star(\sqrt{N_\textrm{config}})$ fit of Fig.~\ref{fig:FigStaticElectrons}f) is pinpointed in dashed red line ($B_\textrm{Z}=3.9$~\si{\milli\tesla}).
}
\end{figure}

\clearpage
\begin{figure}[!h]
\includegraphics[width=\singlecolumn]{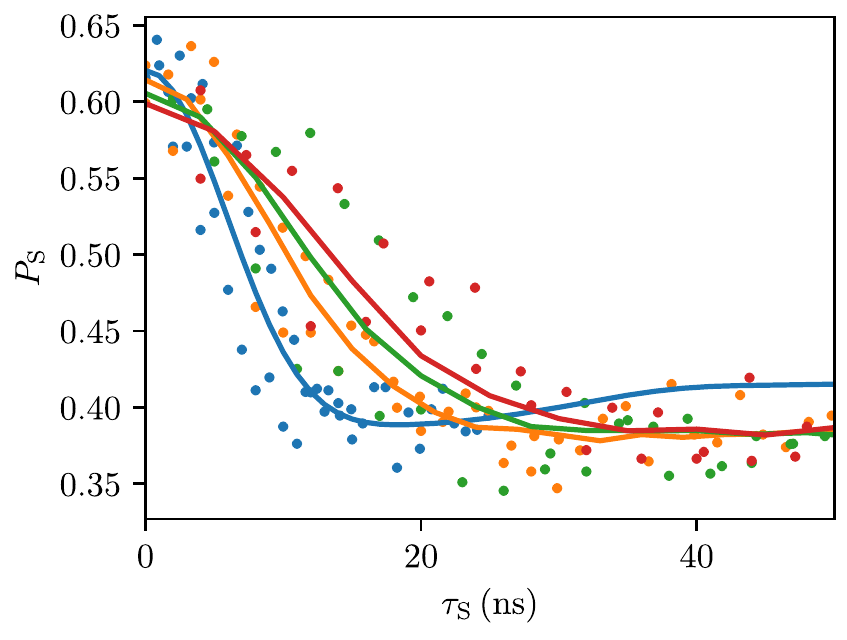}
\caption{
\label{supp:fig:gaussian_quasistatic}
\textbf{Simulated decays of the quasi-static electron spin coherence.}
Same dataset as in Fig.~\ref{fig:FigStaticElectrons}e.
The experimental data (dots) are compared with the simulations (solid lines) of the coherence spin in the quasi-static case (remanence parameter is irrelevant here as each configuration is explored only a single time).
The overshoot in singlet probability described in \cite{Merkulov2002} is observed in the simulations only for the double quantum dot case (blue), and is not observed in the experimental data.
For the two electrons in three (orange), four (green), or five (red) dots, the simulated decays are Gaussian-like.
}
\end{figure}


\clearpage
\begin{figure}[!h]
\includegraphics[width=\doublecolumn]{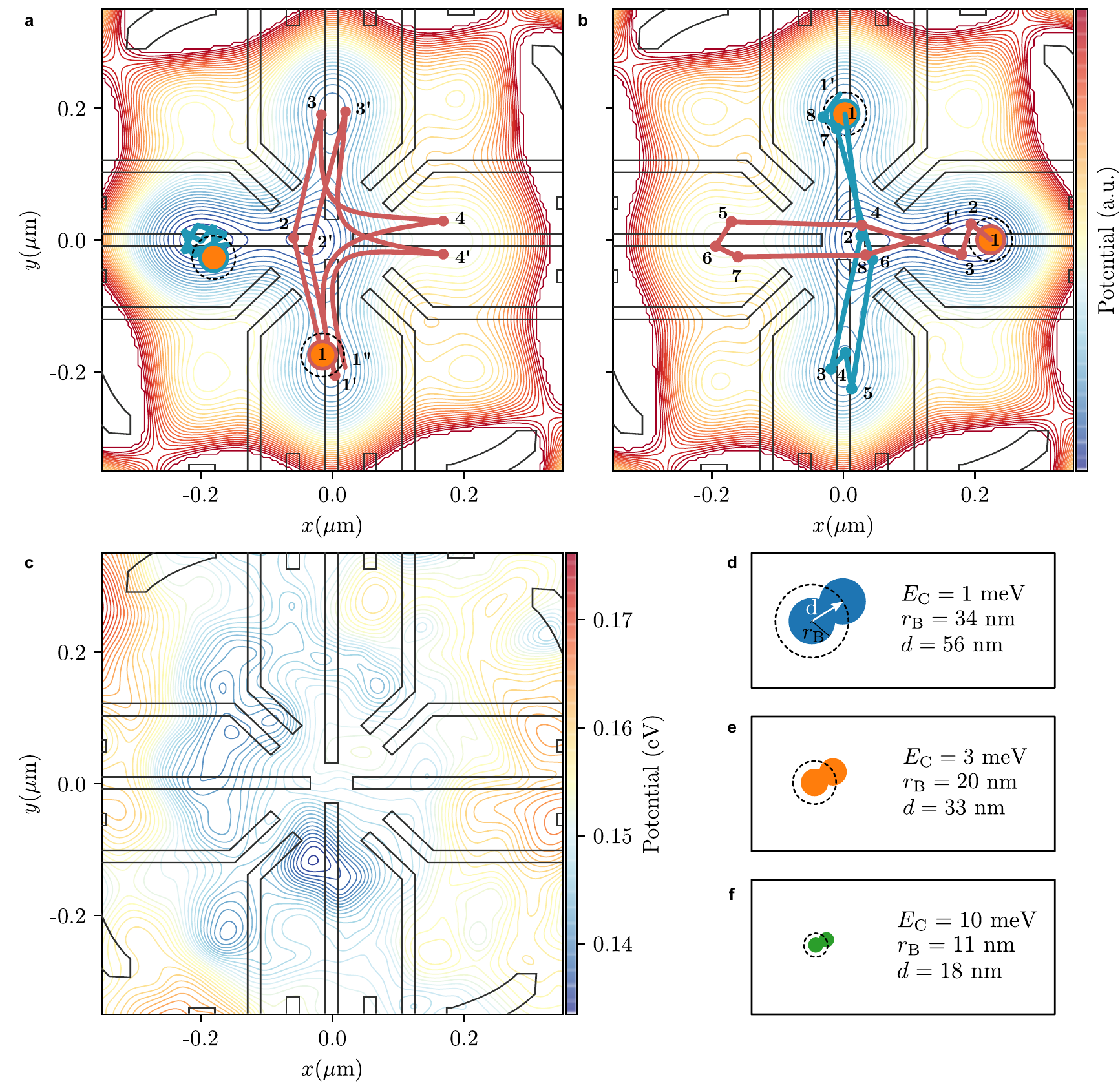}
\caption{
\label{supp:fig:shallowpotential}
\textbf{Electrostatic potential simulations.}
\textbf{a, b,}~Electrostatic potential simulations performed with a gate voltage configurations corresponding to a confinement of two electrons (\textbf{a}) in L-B, and (\textbf{b}) in T-R.
Hypothetical electron trajectories (dots) are overlapped for the two displacement patterns considered.
The solid lines are guides for the eye.
\textbf{a,}~A single electron is periodically displaced between dots (B, C, T, and R), the other electron remains in the L dot.
Four dot configurations are explored, but the dot positions evolve at each period (two periods are drawn, e.g. $1\rightarrow1'\rightarrow1"$).
\textbf{b,}~Two electrons are displaced, one between (L, C, R) and the other between (T, C, B).
Eight dot configurations are explored and here again, the dot positions evolve at each displacement period (a single period is drawn, e.g. $1\rightarrow1'$).
The electrostatic potential simulations were performed by solving the Laplace equation \cite{Bautze2014}.
The orange circles have a radius of 20~\si{\nano\meter} corresponding to the Bohr radius of an electron with a charging energy $E_\mathrm{C}$ of 3~\si{\milli\electronvolt}.
The dashed circles have a radius of 33~\si{\nano\meter}, value computed to reach a remanence $\mathcal{R}=0.5$.
\textbf{c,}~Electrostatic potential simulation of a random donor distribution in the Si-doped layer of the heterostructure (Methods).
The induced potential can localise the dots in different spatial positions than the one intended by the gate geometry.
\textbf{d-f,}~Schematic representations of the Bohr radii and corresponding dot displacements to reach a remanence $\mathcal{R}=0.5$ for different electron charging energies.
}
\end{figure}


\end{document}